# A Probabilistic Upper Bound on Differential Entropy

Joseph DeStefano, *Member, IEEE,* and Erik Learned-Miller

*Abstract*— A novel, non-trivial, probabilistic upper bound on the entropy of an unknown one-dimensional distribution, given the support of the distribution and a sample from that distribution, is presented. No knowledge beyond the support of the unknown distribution is required, nor is the distribution required to have a density. Previous distribution-free bounds on the cumulative distribution function of a random variable given a sample of that variable are used to construct the bound. A simple, fast, and intuitive algorithm for computing the entropy bound from a sample is provided.

## I. INTRODUCTION

The differential entropy of a distribution [6] is a quantity employed ubiquitously in communications, statistical learning, physics, and many other fields. If $X$ is a one-dimensional random variable with absolutely continuous distribution $F(x)$ and density $f(x)$, the differential entropy of $X$ is defined to be

$$H(X) = -\int_{-\infty}^{\infty} f(x) \log f(x) \, dx. \qquad (1)$$

For our purposes, the distribution need not have a density. If there are discontinuities in the distribution function $F(x)$, then no density exists and the entropy is $-\infty$.

It is well known [2] that the entropy of a distribution with support $[y_L, y_R]$ is at most $\log(y_R - y_L)$, which is the entropy of the distribution that is uniform over the support. Given a sample of size $n$ from an unknown distribution with this support, we cannot rule out with certainty the possibility that this sample came from the uniform distribution over this interval. Thus, we cannot hope to improve a deterministic upper bound on the entropy over such an interval when nothing more is known about the distribution.

However, given a sample from an unknown distribution, we can say that it is *unlikely* to have come from a distribution with entropy greater than some value. In this paper, we formalize this notion and give a specific, probabilistic upper bound for the entropy of an unknown distribution using both the support of the distribution and a sample of this distribution. To our knowledge, this is the first non-trivial upper bound on differential entropy which incorporates information from a sample and can be applied to any one-dimensional probability distribution.

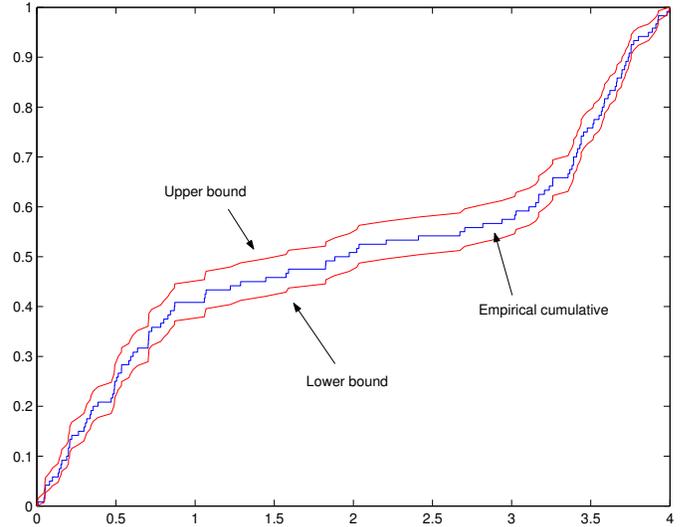

Fig. 1. This figure shows a typical empirical cumulative distribution (blue). The red lines show us the probabilistic bounds provided by Equation 2 on the true cumulative.

## II. THE BOUND

Given $n$ samples,[1] $x_1$ through $x_n$, from an unknown distribution $F(x)$, we seek a bound for the entropy of $F$. Our approach will primarily be concerned with the order statistics[2] of that sample, $y_1$ through $y_n$. We assume that the distribution has finite support and that we know this support. For ease of exposition, we label the left support $y_0$ and the right support $y_{n+1}$ making the support values act like additional order statistics of the sample. But this is done merely for notational convenience and does not imply in any way that these are real samples of the random variable.

We start with a bound due to Dvoretzky, Kiefer, and Wolfowitz [4] on the supremum of the distance between the empirical $n$-sample cumulative, $F_n(x)$, and the true distribution:

$$P(\sup_x |F(x) - F_n(x)| > \epsilon) \le 2e^{-2n\epsilon^2} \equiv \alpha. \qquad (2)$$

Thus, with probability *at least* $\alpha$, the true cumulative does not differ from the empirical cumulative by more than $\epsilon$. This is a distribution-free bound. That is, it is valid for any one-dimensional probability distribution. For background on such bounds and their uses, see [3].

---

[1] We will assume for the remainder of the paper that $n \ge 3$, as this will simplify certain analyses.

[2] The order statistics $y_1, y_2, ..., y_n$ of a sample $x_1, x_2, ..., x_n$ are simply the values in the sample arranged in non-decreasing order. Hence, $y_1$ is the minimum sample value, $y_2$ the next largest value, and so on.

There is a family of cumulative distribution curves $\mathcal{C}$ which fit the constraints of this bound, and with probability at least $\alpha$, the true cumulative must be one of these curves. If we can find the curve in $\mathcal{C}$ with maximum entropy and compute its entropy, then we have confidence at least $\alpha$ that the true entropy is less than or equal to the entropy of this maximum entropy distribution.

Figure 1 shows the relationship between the empirical cumulative and the bounds provided by (2). The piecewise constant blue curve is the empirical cumulative distribution based on the sample. The red stars show the upper and lower bounds on the true cumulative for some particular $\epsilon$. The green curve shows one possibility for the true cumulative distribution that satisfies the probabilisitic constraints of Equation 2. Our goal is to find, of all cumulative distributions which satisfy these constraints the one with the greatest entropy, and then to calculate the entropy of this maximum entropy distribution.

Note that in figure 1, as in all the figures presented, a very small sample size is used to keep the figures legible. As can be seen from equation 3, below, this results in a loose bound. In practice, a larger value for $n$ drives the bound much closer to the emirical cumulative (with $\epsilon$ tending to zero), with a corresponding improvement in the resulting entropy bound.

For a desired confidence level $\alpha$, we can compute a corresponding $\epsilon$ from Eq. 2 that meets that level:

$$\epsilon = \sqrt{-\frac{\ln \frac{1-\alpha}{2}}{2n}}. \qquad (3)$$

We conclude that with probability $\alpha$, the true distribution lies within this $\epsilon$ of the empirical distribution at all $x$.

### A. Pegs

The points $u_i = (y_i, \frac{i}{n+1} + \epsilon)$ (resp. $l_i = (y_i, \frac{i}{n+1} - \epsilon)$) describe a piecewise linear function $F_u$ (resp., $F_l$) marking the probabilistic upper (resp., lower) boundary of the true cumulative $F$. We call these points the upper (resp. lower) *pegs*, and note that we clip them to the range $[0, 1]$. Also, our knowledge of the support of the distribution allows us to take $u_0 = l_0 = (y_0, 0)$ and $u_{n+1} = l_{n+1} = (y_{n+1}, 1)$.

The largest of the entropies of all distributions that fall within $\epsilon$ of the empirical cumulative (i.e., entirely between $F_u$ and $F_l$) provides our desired upper bound on the entropy of the true distribution, again with probability $\alpha$. The distribution that achieves this entropy we call $F_M$.

To illustrate what such a distribution looks like, one can imagine a loose string threaded between each pair of pegs placed at the $u_i$ and $l_i$, as in Figure 2. When the ends are pulled tight, the string traces out a distribution which, as we show below, has the greatest possible entropy of all such distributions. Since this distribution turns out to be piecewise linear, its entropy, which again is the probability-$\alpha$ bound on the true entropy, is easily computed.

It is interesting to note that extreme values of $\epsilon$ correspond to the empirical distribution itself ($\epsilon = 0$), and to the naïve entropy estimate of $\log(y_n - y_1)$ ($\epsilon = 1$), so in some sense our algorithm produces a bound between these two extremes.

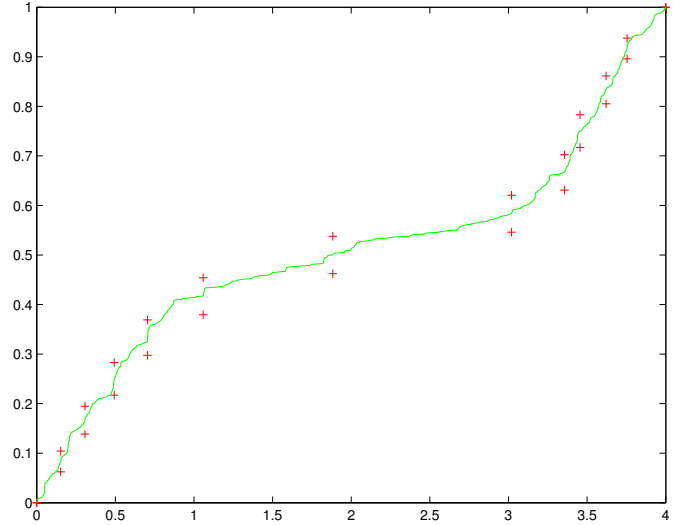

Fig. 2. Pulling tight the ends of a loose string (the green line) threaded through the pegs will cause the string to trace out the maximum-entropy distribution that satisfies the bound (2).

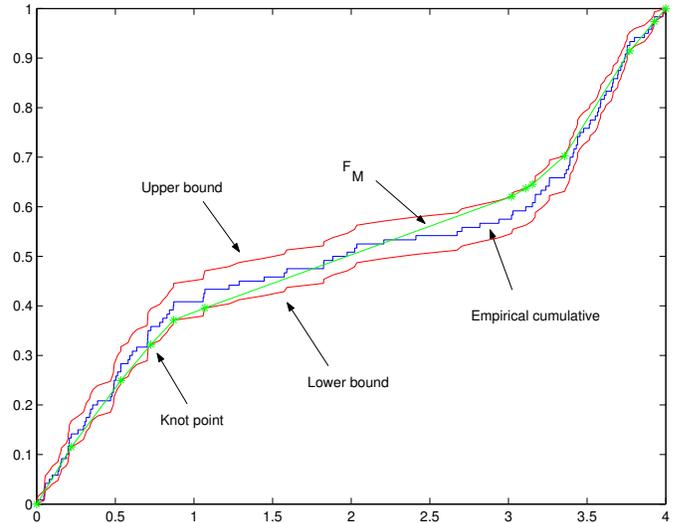

Fig. 3. This figure shows the maximum entropy cumulative distribution which fits the constraints of the Dvoretzky-Kiefer-Wolfowitz inequality for the given empirical cumulative distribution. Notice that the cumulative is piecewise linear, implying a piecewise constant density function. With probability at least $\alpha$, the true cumulative distribution $F$ has entropy less than or equal to this maximum entropy distribution.

## III. THE STRING-TIGHTENING ALGORITHM

We first develop several properties of $F_M$ that provide the basis for our algorithm, which we call the string-tightening algorithm.

*Lemma 1:* $F_M$ is linear between consecutive order statistics.

*Proof:* Although the true distribution may have discontinuities, the entropy of any such distibution is $-\infty$. We therefore can restrict our search for $F_M$ to those distributions with densities.

First we must show that the bound (2) allows $F_M$ to be linear. That is, we must show that the curve between two consecutive order statistics is not restricted by the bound so that it cannot be linear. To do this, it suffices to show that the bound $\epsilon$ is always larger than the step size $\frac{1}{n}$ in the cumulative distribution. The bound is minimum when $\alpha = 0$, so we have

$$\begin{aligned} \epsilon &\geq \sqrt{-\frac{\ln\frac{1-0}{2}}{2n}} \\ &\approx \frac{.5887}{\sqrt{n}} \\ &> \frac{1}{n}, \forall n \geq 3. \end{aligned}$$

Given that it *can* be linear, we next show that it *must* be linear between consecutive order statistics. The proof is by contradiction. Suppose that between two consecutive order statistics, $y_i$ and $y_{i+1}$, $F_M$ is not linear.

Note that the entropy function is separable into integrals over the region of interest $[y_i, y_{i+1}]$ and the remainder of the real line $\overline{[y_i, y_{i+1}]}$:

$$\begin{aligned} H(F_M) = &- \int_{y_i}^{y_{i+1}} f_M(x)\log f_M(x)dx \\ &- \int_{\overline{[y_i,y_{i+1}]}} f_M(x)\log f_M(x)dx. \end{aligned}$$

Because of this separability, conditioned on specific values for $f_M(y_i)$ and $f_M(y_{i+1})$, $f_M$ must maximize each of the terms above separately.

Let $\mathcal{F}$ be the set of all monotonic non-decreasing functions over $[y_i, y_{i+1}]$ such that if $f \in \mathcal{F}$, then $f(y_i) = f_M(y_i)$ and $f(y_{i+1}) = f_M(y_{i+1})$. Also, let $C = f_M(y_{i+1}) - f_M(y_i)$. Then

$$\arg\max_{f \in \mathcal{F}} - \int_{y_i}^{y_{i+1}} f(x)\log f(x)dx \tag{4}$$

$$= \arg\max_{f \in \mathcal{F}} - \int_{y_i}^{y_{i+1}} f(x)\left[\log f(x) - \log(C)\right]dx \tag{5}$$

$$= \arg\max_{f \in \mathcal{F}} - \int_{y_i}^{y_{i+1}} f(x)\log\frac{f(x)}{C}dx \tag{6}$$

$$= \arg\max_{f \in \mathcal{F}} - \int_{y_i}^{y_{i+1}} \frac{f(x)}{C}\log\frac{f(x)}{C}dx. \tag{7}$$

.

The last expression is just the entropy of the distribution $g(x) = \frac{f(x)}{C}$, which is a properly normalized probability distribution over $[y_i, y_{i+1}]$. It is well-known [2] that $g(x)$ must be uniform to maximize entropy over a finite interval. This in turn, implies that $f(x)$ must be uniform to maximize (4). Hence, if $F_M$ is not linear between $y_i$ and $y_{i+1}$ then it cannot be the entropy maximizing distribution, contradicting our assumption. ∎

Thus $F_M$ is piecewise linear, with any "bends", or changes in slope, occuring only at the sample points. Intuitively, as the string is tightened, it is clear that these slope changes can occur only at the pegs, which we formalize here.

*Lemma 2:* An increase (decrease) in slope of $F_M$ can occur only at the upper (lower) peg of a sample.

*Proof:* By contradiction. Suppose that there are two connected segments of $F_M$, $(y_{i-1}, a) - (y_i, b)$ and $(y_i, b) - (y_{i+1}, c)$, with $u_i > b$ and $b$ below the segment $(y_{i-1}, a) - (y_{i+1}, c)$ (i.e., the slope increases at $(y_i, b)$). Then there is an interval $[y_i - \delta, y_i + \delta]$, $\delta > 0$ where the line segment $F_M(y_i - \delta) - F_M(y_i + \delta)$ lies entirely between $F_l$ and $F_u$. The argument of lemma 1 shows that this segment maximizes the entropy on $[y_i - \delta, y_i + \delta]$, and thus $F_M$, being maximal, cannot pass through $(y_i, b)$, contradicting the assumption. A similar argument applies for a decrease in slope. ∎

Thus $F_M$ is completely described by the sequence of pegs that it touches, which we call the *knot points*. The string-tightening algorithm is a left-to-right search for knot points, starting with the known first knot, $l_0$, as follows.

Given a knot $K$ (except the last knot, $u_{n+1}$), we define a *reachable peg* as any peg $P$ to the right of $K$ for which the segment $\overline{KP}$ is contained between $F_l$ and $F_u$. The *candidate upper (lower) peg* is the rightmost upper (lower) reachable peg (i.e., the one with highest index). Lemma 2 ensures that one of these two candidates must be the next knot to follow $K$. If the upper candidate is $u_{n+1}$, it is added as the final knot. Otherwise, to determine which is the knot, consider all the pegs to the right of the two candidates. As shown in figure 4, the rays from the knot to the upper and lower candidate pegs define a wedge, and for each sample $y_i$ to the right of the candidates, neither of the pair of pegs $u_i$ or $l_i$ can lie within that wedge (by the definition of the candidates as the rightmost such pegs).[3] Since $u_{n+1} = l_{n+1}$, there must be a at least one pair of pegs that are both on one side of the wedge. Let $j$ be the smallest index of all such pairs. If $l_j$ is above the wedge, then the slope of $F_M$ must increase after passing the upper candidate in order to pass between $l_j$ and $u_j$; thus by Lemma 2 the candidate upper peg is the next knot. Otherwise, the candidate lower peg is the next knot.

This process is repeated until the final knot, $u_{n+1}$ is added. The segments connecting the knots form $F_M$. Its entropy, which is our probability-$\alpha$ bound on the entropy of the true distribution $F$, is easily computed from the spacings of the knots and the slopes of the segments between them, since $f_M$ is piecewise constant. Writing the knot $K_i$ as $(a_i, b_i)$, we have

$$H_M = -\sum(b_{i+1} - b_i)\log\frac{b_{i+1} - b_i}{a_{i+1} - a_i}. \tag{8}$$

### A. Performance

Figure 5 compares the 95% confidence bound produced by the string-tightening algorithm for several distributions (shown in figure 6) to the true entropy, obtained by numerical integration. The naïve bound $\log(y_n - y_1)$ is also plotted at the top of each graph.

## IV. A TIGHTER BOUND

The bound on the distribution provided by (2) allows for the same degree of uncertainty at all points. Intuitively, it seems we should be able to bound the distribution more tightly near the ends of the support than in the middle. For empirical support of this intuition, we generated 10000 experiments with 100 samples each from a known distribution, and recorded which of the order statistics were outside the bound (2) for $\alpha = 0.95$. The histogram of this data in figure 7 clearly shows that the bound provided by (2) is not as tight as it could be near the ends of the distribution: One would expect that a bound that was as tight as possible everywhere would miss equally often at all points.

---

[3]In general, the upper and lower candidates can have different indices, i.e., one of them may be further "right" than the other.

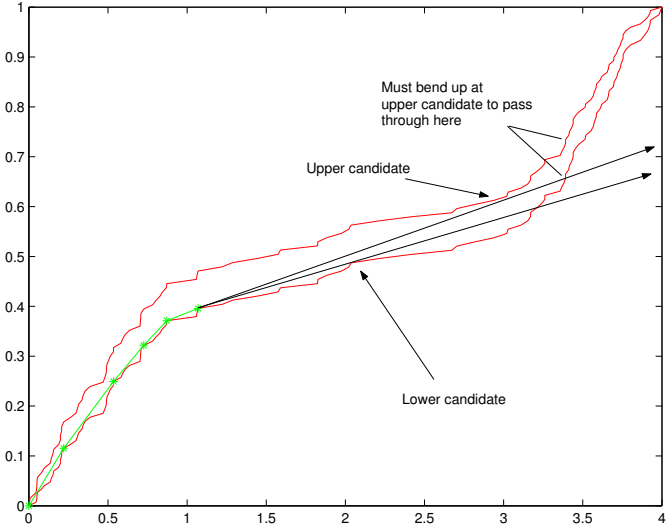

Fig. 4. The upper and lower candidate pegs define a wedge. The first subsequent pair of pegs that are both on one side of the wedge determine whether $F_M$ must bend down at the lower candidate, or bend up at the upper candidate.

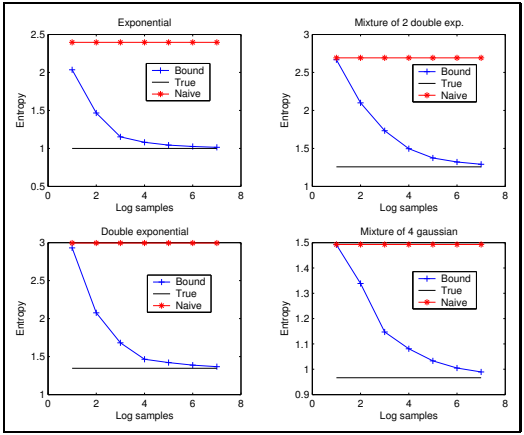

Fig. 5. The 95% confidence bound quickly becomes much tighter than the naïve bound.

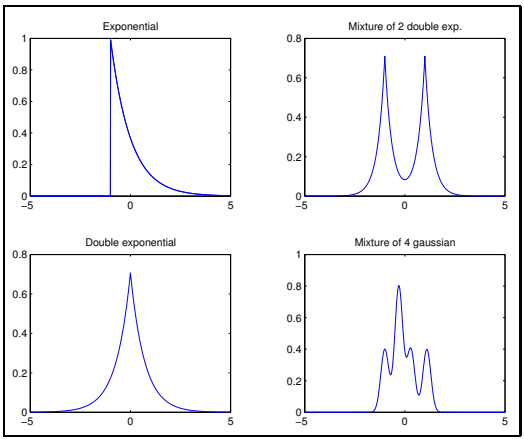

Fig. 6. The four distributions used in the comparisons in figure 5.

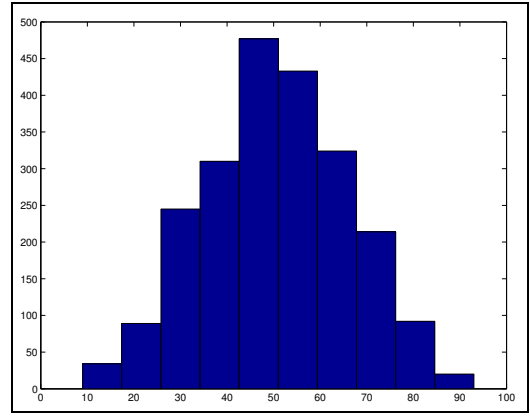

Fig. 7. Bound (2) is too loose near the edges.

To tighten the bound, we use the fact that for samples $x_i$ from a distribution $F(x)$, the values $F(x_i)$ are uniformly distributed on $[0, 1]$ [5]. Therefore the value of $F(y_i)$ has the same distribution as the $i$-th order statistic of a uniformly distributed sample, i.e., it is beta distributed with paramters $i$ and $n - i + 1$ [1]. Its mean is $\frac{i}{n+1}$, or $F_n(y_i)$. One could, in principle, determine intervals $(a_i, b_i)$ such that

$$P(\forall i \; F(y_i) \in (a_i, b_i)) = \alpha \qquad (9)$$

for a given $\alpha$, and for which

$$\forall i \; P(F(y_i) \notin (a_i, b_i)) = Q$$

for some constant $Q$. Using these as the pegs (i.e., taking $l_i = a_i$ and $u_i = b_i$) would then flatten out the histogram.

Given the combinatoric nature of the joint distribution of order statistics, an exact computation of these bounds is likely to be intractable. We instead determine useful intervals empirically by choosing a value of $Q$, selecting bounds based on the beta distribution of each order statistic, and verifying that the fraction of a large number of randomly generated samples that fall entirely within those bounds is at least $\alpha$.

Were the order statistics independent, one would have to choose an interval that gave $P(F(y_i) \in (a_i, b_i)) = \alpha^{\frac{1}{n}}$ in order to satisfy equation (9). However, since they are strongly dependent, a much smaller value will suffice. We found that using intervals such that $P(F(y_i) \in (a_i, b_i)) = 0.999$ produced an effective probability $\alpha = 0.97$ that all the order statistics are within their intervals simultaneously. The nature of the beta-distributions of each statistic results in these intervals being very narrow near the edges, and very close to $\epsilon$ (from (2)) near the middle. Figure 8 shows this bound as the

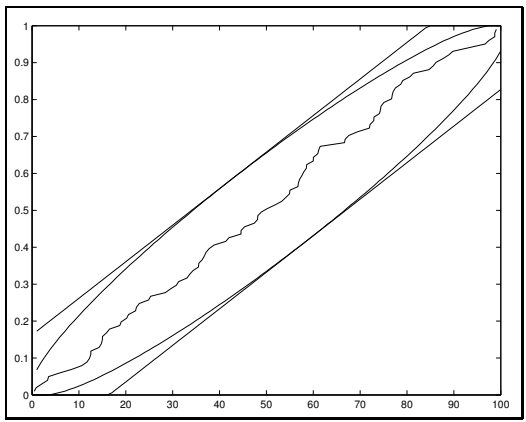

Fig. 8. Empirically generated bounds are tighter than those given by the bound (2). The jagged line is one of the many possible cumulatives that satisfy the bounds.

curved lines, with the straight lines giving the bound given by $\epsilon$.[4]

## V. Conclusion

We have shown how distribution-free bounds on the cumulative distributions of unknown one-dimensional probability densities can be used to give sample-based probabilistic bounds on the entropies of distributions with known support. As an alternative to providing the support of the distribution, one can provide bounds on the mean log probability density of the tails of a distribution, and still provide similar bounds. We leave this topic to future work.

We have provided a simple algorithm to compute this bound exactly from samples taken from the unknown distribution. A by-product of the algorithm is an explicit representation of $F_M$, the distribution that achieves the computed bound. The simple form of $F_M$ makes it convenient for use in resampling applications.

[4]The "unknown" distribution in this example is a uniform distribution, for clarity. Similar differences are found near the support boundaries for any distribution.